\documentclass[aps,prd,onecolumn,showpacs,preprintnumbers,amsmath,amssymb]{revtex4}
\usepackage{graphicx}
\usepackage{dcolumn}
\usepackage{bm}
\usepackage{epsfig}
\begin{document}
\title{Reply to ``{C}omment on `{T}est of constancy of speed of light with rotating cryogenic optical resonators' ''}

\author{P. Antonini$^{1}$, M. Okhapkin$^{1}$, E. G\"okl\"u$^{2}$, S. Schiller$^{1}$}

\affiliation{1: Institut f{\"u}r Experimentalphysik, Heinrich-Heine-Universit{\"a}t D{\"u}sseldorf,
40225 D{\"u}sseldorf, Germany\\
2: ZARM, University of Bremen, 28359 Bremen, Germany}

\vskip -.5in
\date{\today}
\begin{abstract}
An improved analysis of the Michelson-Morley-type experiment by P.
Antonini et al. (Phys. Rev. A. 71, 050101 (2005)) yields the
Robertson-Mansouri-Sexl theory parameter combination
$\beta-\delta-1/2=(-0.6\pm 2.1\pm 1.2)\cdot10^{-10}$
 and the Standard Model Extension theory parameter $(\tilde{\kappa}_{e-})^{ZZ} = (-2.9\pm 2.2)\cdot 10^{-14}$.
\end{abstract}
 \maketitle

\def\kappaezz{(\tilde{\kappa}_{e-})^{ZZ}}

1.) As stated in our paper \cite{Antonini2005} on page 050101-4, the measured value of $\kappaezz$ was consistent with a systematic effect due to variation of the
temperature affecting the experiment and therefore we gave its value as a probable upper limit   \cite{error}.

In order to characterize the systematic effects more precisely, we have analyzed in detail
the correlations between various monitored parameters and the beat frequency. The monitored parameters were (apart from the two tilt angles)
 two laboratory temperatures measured
at the top and bottom of the cryostat, and five temperatures at different positions inside the cryostat.

A linear regression analysis showed that
there are strong correlations between these parameters and the beat frequency. The data was therefore decorrelated, whereby
time intervals of 10000\,s were decorrelated individually, in order to allow
for changing environmental conditions.
This removed to a large extent variations of the beat frequency with respect to a nearly constant drift.

The isotropy violation analysis was repeated for the decorrelated
data. For the rotations considered in \cite{Antonini2005} the
average is $\langle 2C\nu\rangle=-2.4\,$Hz with a sample standard
deviation of 1.9\,Hz, and $\langle 2B\nu\rangle=0.8\,$Hz, with a
sample standard deviation 2.6\,Hz. A large number of rotations
performed under different experimental conditions allows us to
estimate the uncertainty due to (identifiable) systematic effects at
1.8\,Hz. The result is $\kappaezz=-2.9\cdot10^{-14}$, with an error
of $\pm2.2\cdot10^{-14}$, dominated by the systematic effects
\cite{Proceedings}.

\vskip .1in

We now discuss the arguments given in the comment by Tobar et al.
\cite{tob05a}. Giving a final result with an error equal to the
standard deviation of the data, as Tobar et al have done, is
certainly both a simple and a conservative estimate, but the issue
is whether this estimate is too conservative. If one believes that
the systematics are dominant (as may be safely assumed for the
experiments discussed here)
 then this statement is simply a statement of the instability of the apparatus.
Even quoting such an error, there is no certainty that the confidence interval
includes the true value, since a partial cancellation between an unknown {\em constant} systematic and the signal is still possible. An experiment
can never disprove this possibility.

Tobar et al. state that the systematic signal will cancel if the phase is random.
If that is the case, then a correct statement of the standard error is the standard deviation of the sample mean. We disagree with their
statement that it should be the sample standard deviation. In practice it may be difficult to asses how well the systematic averages to zero, because usually it cannot be clearly identified.
Moreover, a  very small number $n$ of data points (as in the work by Stanwix et al. where $n=5$ \cite{Stanwix2005}) makes the calculated mean
(in other words, a good cancellation of the systematic due to random phase) even more uncertain, so that quoting the standard deviation as the final error can be accepted.

We emphasize that
bounds provided by a single experiment may still be affected by the presence of a (constant)  systematic effect.
However,  the results of the three recent experiments \cite{Antonini2005,Proceedings,Stanwix2005,Herrmann2005} are consistent with each other.
Because they were performed by independent groups with different techniques, it is not likely that they all exhibit a strong cancellation between
 the respective systematics and a substantial nonzero value of $\kappaezz$.

2.) Concerning our statement of the improvement achieved  by our
experiment compared to previous ones, Tobar et al assume that we
quoted the improvement in the uncertainties of the experiment.
However, we did not. In order to quantify the progress provided by a
experiment by a single number we consider more useful to quote the
improvement in the stringency of the test of the acutal hypothesis
under study, namely that the speed of light is isotropic. The
previous experiments \cite{Mueller2003,Wolf2004} were consistent (at
the 90\% confidence level) with isotropy at the level of
$|\beta-\delta-1/2|<4.1\times10^{-9}$, and $4.2\times10^{-9}$,
respectively, while our reported result implied a bound of
$5.1\times10^{-10}$. The latter bound is more stringent by a factor
8, which replaces our imprecise statement "about a factor 10" given
in the paper.

For better comparison with the work by Stanwix et al.\,\cite{Stanwix2005} and Herrmann et al.\,\cite{Herrmann2005},
we have analyzed in a similar way a larger number of rotations (940, grouped in several sets) according to the RMS model.
Our result is
$\beta-\delta-1/2=(-0.6\pm 2.1\pm1.2)\cdot10^{-10}$, where the first uncertainty is statistical and the second is systematic, due to uncertainties in
tilt sensitivities and due to laser power variations \cite{Proceedings}.

\end{document}